\documentclass[floatfix,10pt,onecolumn,showpacs,amsmath,amssymb]{revtex4}

\usepackage{epsf}
\usepackage{graphicx}  
\usepackage{dcolumn}   
\usepackage{bm}        

\newcommand{\be}{\begin{equation}}
\newcommand{\en}{\end{equation}}
 \newcommand{\bea}{\begin{eqnarray}}
 \newcommand{\ena}{\end{eqnarray}}
  \newcommand{\sch}{Schwarzschild}

\begin{document}

\title{Black holes and gravitational waves in three-dimensional f(R) gravity}
\author{Hongsheng Zhang$^{1,2~}$\footnote{Electronic address: hongsheng@shnu.edu.cn}, Dao-Jun Liu $^{1,2~}$ \footnote{Electronic address: djliu@shnu.edu.cn},  Xin-Zhou Li$^1$ \footnote{Electronic address: kychz@shnu.edu.cn} }
\affiliation{ $^1$Center for
Astrophysics, Shanghai Normal University, 100 Guilin Road,
Shanghai 200234, China\\
$^2$State Key Laboratory of Theoretical Physics, Institute of Theoretical Physics, Chinese Academy of Sciences, Beijing, 100190, China
}

\date{ \today}

\begin{abstract}
 In the three-dimensional pure Einstein gravity, the geometries of the vacuum space-times are always
trivial, and gravitational waves (gravitons) are strictly forbidden. For the first time, we find a vacuum
circularly symmetric black hole with nontrivial geometries in $f(R)$ gravity theory, in which a true
singularity appears. In this frame with nontrivial geometry, a perturbative gravitational wave does exist.
Beyond the perturbative level, we make a constructive proof of the existence of a gravitational wave in $f(R)$
gravity, where the Birkhoff-like theorem becomes invalid. We find two classes of exact solutions of
circularly symmetric pure gravitational wave radiation and absorption.

\end{abstract}


\preprint{arXiv: }
 \maketitle

   \section{Introduction}
   Pure gravity in three-dimensional space-time has been extensively studied over many aspects, for a review see \cite{carlip}. But its status is still unsettled. It is well-known that the three-dimensional pure gravity is trivial, in the sense that the manifold must be a maximally symmetric space locally, and thus gravitational wave is forbidden. That is to say, the three-dimensional gravity becomes a topological theory: the differences among different configurations in this theory only arose when we consider the boundary effects. A very simple and intuitionistic explanation of this point is shown as follows. The Weyl tensor always vanishes in three-dimensional space-time, and thus the Riemann tensor can be expressed in Ricci tensor. From Einstein equation the Ricci tensor is proportional to the metric in pure gravity. Therefore, the Riemann tensor is completely determined by the metric (without the derivative terms of the metric), which defines a maximally symmetric space-time. There are only a few non-propagating configurations, just like the solutions of time independent Schr\"{o}dinger equation in quantum mechanics, can exist in this frame \cite{deser}. Clearly, a graviton does not exist since there is no space-time ripple that can emerge and propagate, which implies that the quantum theory of three-dimensional gravity may be of absence.

  The arguments of non-existence of quantum theory in three-dimensional gravity are impaired by the equivalence of three-dimensional gravity and Chern-Simons gauge theory. The Chern-Simons quantum gauge theory does exist, and furthermore, is renormalizable. So one expects that a well-defined quantum theory of the three-dimensional gravity also exists. It is worth noting that the equivalence between the three-dimensional gravity and Chern-Simons theory is constrained in local classical and perturbative sense. They have several important differences when one observes their global non-perturbative properties. For example, in the gravity aspect, it is natural to sum over all possible topologies of the space-time manifold. But in the aspect of Chern-Simons gauge formulation, it is difficult to find the necessity to do so. As there are important differences between three-dimensional gravity and Chern-Simons formulation, one may simply give up the efforts to find a quantum theory for three-dimensional gravity. However, the discovery of BTZ black hole forces us to reconsider this problem \cite{btz}. The BTZ black holes have horizon, mass, entropy, and angular momentum, though its local geometry is just anti-de Sitter. Especially, these physical quantities obey the thermodynamic laws, like a usual four-dimensional black hole. A significant discovery in 1970's is that gravity is inherently related to thermodynamics. Several  deep and remarkable approaches have been made in gravi-thermodynamics, for example the black radiation \cite{Hawking}. Recently we find that the spherically symmetric metrics can be derived directly from thermodynamic considerations \cite{zhang}. Generally the macroscopic thermodynamic properties have microscopic statistical origins. The existence an entropy proportional to the perimeter of the horizon hints that there are rich freedoms in the BTZ space-time.  Thus it is natural to explore whether we can find propagating modes by introducing some new freedoms into the theory.

                         The propagation freedoms in three dimensional (modified) gravity have been investigated.   Some interesting points of this problem can be found in \cite{carlip, witten}. Generally, modified gravity permits extra freedoms than general relativity, for example the dilaton field in scalar-tensor gravity. The additional freedoms in modified gravity may lead non-trivial geometries in 3-dimensions. A topological massive gravity is introduced in \cite{deser2}, in which a spin-2 graviton with propagation degree appears. The Hilbert-Einstein action together with a special combination of $R^2$-terms is equivalent to a Pauli-Fierz action at the linearized level \cite{bht}. This new massive gravity has been studied in \cite{bt2}. Exact black hole solutions with non-trivial geometries have been found in this massive gravity frame in three dimensions \cite{ott}. A parity-preserving model of three-dimensional gravity with propagating torsion freedom in asymptotic AdS space in studied in \cite{Blago}.
 In this paper, we make a different approach towards this problem from the view of a higher derivative gravity theory.  We present an exact solution in a three-dimensional higher derivative gravity as a heuristic example to display a possible way to introduce graviton.

 In the next section we show a black hole solution in $R^{d+1}$ gravity. In section III we present the gravitational wave solution in $R^{d+1}$ gravity, and we conclude this paper in section IV.

     \section{The~ black~ hole~ solution}
         It is wellknown that three-dimensional Einstein gravity does not permit a graviton. The theory is non-renormalizable even if we consider an off-shell graviton, since the gravitational constant has a dimension of length. The studies of the renormalizability of gravity leads to higher derivative theory \cite{stelle}, for some early thoughts on this issue, see \cite{dewitt}. $f(R)$ gravity naturally encodes the essential properties of higher derivative theories, meanwhile it is relatively tractable compared to some more complicated considerations, for a review see \cite{soti}. Once the higher derivative terms are introduced, they will dominate the behaviour of the space-time dynamics at high energy region, such as the regions around the big bang singularity or a black hole singularity. We shall show that a true singularity is possible in three-dimensional pure $f(R)$ gravity, while there is no true singularity in three-dimensional Einstein gravity. Accompanying to the appearance of the true singularity, the local geometry becomes non-trivial, and thus an on-shell graviton (gravitational wave) is also possible.

   We start from an action in $f(R)$ gravity,
   \begin{gather}
   S=\frac{1}{16\pi G }\left(\int_{\cal M} d^{3}x\sqrt{-{\rm det}(g)}~R^{d+1}+\int_{\cal \partial M} d^{2}x\sqrt{-{\rm det}(h)}~2B_o\right),
     \label{action1}
   \end{gather}
%
   where $G$ is the three-dimensional constant with a dimension of length, ${\cal M}$ is the manifold in consideration and ${\cal \partial M}$ denotes its boundary, $g$ labels the metric on
   ${\cal M}$ and $h$ labels its induced metric on ${\cal \partial M}$, $R$ is the Ricci scalar for $g$, and $d$ is a constant.
   With a proper boundary term $B_o$ the corresponding field equation reads,
   \begin{gather}
 (1+d)R^{d}R_{ab }-\frac{1}{2}R^{d+1}g_{ab }
 -(1+d)\nabla _{a }\nabla
 _{b }R^{d}+g_{ab }(1+d)\square R^{d}=0.
 \label{field}
   \end{gather}
%
  A general circularly symmetric metric with 2 Killing fields  reads,
\be
ds^2=-A(r)dt^2+\frac{1}{B(r)}dr^2+r^2d\theta^2,
\label{metric1}
\en
 Then, the $\theta$-$\theta$ component of  (\ref{field}) and its trace read
\be
\label{22}
\left[\frac{\partial^2}{\partial r^2}+\left(\frac{A'}{2A}+\frac{B'}{2B}\right)\left(\frac{\partial}{\partial r}-\frac{1}{r}\right)-\frac{R}{2(1+d)B} \right]R^d=0
\en
and
\be
\label{trace}
\left[\frac{\partial^2}{\partial r^2}+\left(\frac{1}{r}+\frac{A'}{2A}+\frac{B'}{2B}\right)\frac{\partial}{\partial r}-\frac{(1-2d)R}{4(1+d)B} \right]R^d=0,
\en
respectively, where
the Ricci scalar $R$ is given by
\be
\label{ricci}
R=-B\left[\left(\frac{A'}{A}\right)'+\left(\frac{A'}{A}+\frac{B'}{B}\right)\left(\frac{1}{r}+\frac{A'}{2A}\right)\right].
\en
 (\ref{22}) and (\ref{trace}) are closed for the two functions $A(r)$ and $B(r)$. Obviously, $A=C_1$, $B=C_2$ ($C_1$ and $C_2$ are two constants) is a solution with
 \be
 R_{ab}=0,
 \en
 and
 \be
 R=0,
 \en
 for the field equation (\ref{field}). This is a trivial Minkowskian one.

  Besides this Minkowskian solution, we find that there exist non-trivial solutions for this three-dimensional theory.  Although there is no general method to solve such non-linear equations as the field equation (\ref{field})  at fourth-order, we find a proper ansatz under which the analytical forms of $A$ and $B$ can be obtained. Now we introduce the following ansatz,
\be
\label{ransa}
R=-\frac{kL}{r^p},
\en
where $L>0$, and $k=-1,~0,+1$. With this additional ansatz, we find the following solution,
\be
A=Br^{\frac{2pd(1+pd)}{1-pd}},
\label{Ar}
\en
\be
B=k\frac{L(1-d  p) r^{2-p}}{2 (d +1) \left(d ^2 (2 d +1) p^3-p+2\right)} + C r^{\frac{2 p^2 d ^2}{p d -1}},
 \label{Br}
\en
and
\be
p=\frac{1-2d}{(1+2d)d},
\en
 where $C$ is an integration constant. One can check that (\ref{ransa}) are satisfied for $A$ and $B$ in (\ref{Ar}) and (\ref{Br}). The situations of different $k$ are something like the 3 different spatial geometries at $r=$ constant in the case of topological \sch~ solution. However, we emphasize that the local spatial geometries at $r=$ constant are the same for different $k$ in the above solution, because a one-dimensional object can not be intrinsically curved.

 From the ansatz, one explicitly sees that the geometries are non-trivial when $k=\pm 1$. A true singularity appears at $r=0$ for $p>0$, and at $r\to \infty$ when $p<0$. The Ricci scalar (\ref{ransa}) vanishes when $k=0$, but it is still a curved space, which can be seen from the Kretschmann scalar,
 \be
 R_{abcd}R^{abcd}=\frac{C^2(1-2 d)^2 \left(1+12 d^2\right)}{2 d^2 (1+2 d)^2}r^{-\frac{1+12 d^2}{d+2 d^2}}.
 \label{kret}
 \en
 $p=0$ (corresponding to $d=1/2$) is a special point, at which the solution degenerates to a Minkowskian one for all 3 cases of $k$. This means that $R^{3/2}$-gravity permits a Minkowskian state. For a general $k$, (\ref{kret}) becomes
 \bea
 R_{abcd}R^{abcd}=\frac{C^2(1-2 d)^2 \left(1+12 d^2\right)}{2 d^2 (1+2 d)^2}r^{-\frac{1+12 d^2}{d+2 d^2}} &+& \frac{ (1-2 d )^2 }{2 (1-6 d ) d  (d +1)}C kL r^{\frac{8}{2 d +1}-\frac{3}{2 d }-3}\notag\\
  &+& \frac{24 d ^4+40 d ^3+10 d ^2-8 d +1}{36 d ^4+60 d ^3+13 d ^2-10 d +1}L^2k^2 r^{\frac{8}{2 d +1}-\frac{2}{d }}.
  \ena
 It is clear that the the Kretschmann scalar is divergent when $d\to 0$, which implies that this solution is a nonperturbative one compared to the trivial Minkowskian solution. This property also can be understood by our ansatz. $R$ is not a continuous function when $d=0$ ($p\to \infty$): It is a step (the step goes to infinity) function at $r=1$. Hence the metric is not regular either at the point $d=0$ in the parameter space.

 The true singularity is enclosed by an apparent horizon $l(r)=0$, which satisfies,
 \be
 h^{ab}\partial_al\partial_bl=0,
 \en
 where $h$ is the induced metric,
 \be
 h=-A(r)dt^2+\frac{1}{B(r)}dr^2.
 \en
The apparent horizon dwells at,
 \be
 r_a=\left(-\frac{ C (6 d -1) (d +1)}{d  ( d +1/2) kL}\right)^{\frac{2 d }{6 d -1}}.
   \en
 A physical horizon requires,
 \be
 C (6 d -1) (d +1)d  ( d +1/2) k<0.
 \en
  For the case with $k=0$, a horizon cannot appear and generally the singularity becomes a naked one.

 \section{Gravitational wave}
 Since the geometries of the three-dimensional spacetime can be non-trivial, naturally the space-time ripples become possible. And thus we can discuss the perturbative gravitational wave  in the background of the black holes.  Beyond perturbative level, an exact solution says the final words of the existence of gravitational wave. We find that the following metric,
 \be
  ds^2=-\overline{A}(r,u)du^2 \pm {2}{\overline{B}(r)}dudr+r^2d\theta^2,
  \label{dynametr}
  \en
 solves   (\ref{field}), where
 \be
 \overline{A}(r,u)=\frac{1}{2 r}\left(\frac{d (1+2 d) kL r^3}{6 d^2 +5 d-1}+2 r^{\frac{1}{2 d}} {\cal G}(u)\right),
 \en
 and
 \be
 \overline{B}(r)=r^{\frac{1/2-d}{d(1+2d)}}.
 \en
  Here ${\cal G} (u)$ is an arbitrary $C^4$-function of $u$. The source of the above metric reads,
  \be
  T=\mp\frac{|k| d} {4\pi(1+2 d)} r^{-\frac{4d}{1+2 d}} \frac{ d{\cal G}}{du} du\otimes du,
  \label{source}
  \en
  which is independent of $L$.
  It is a null source that satisfies the condition
  \be
  g^{ab}T_{ab}=0.
  \en
  However, it is not a three-dimensional electromagnetic field, since three-dimensional electromagnetic field has a non-zero trace. The cases  $k=\pm 1$ describe  dynamical space-times with sources (\ref{source}). Surely, gravitational waves propagate in these space-times.

   (\ref{dynametr}) is closely related to the black hole solution in the last section. $\cal G$=constant is a degenerated case, which is just (\ref{metric1}) in a different coordinates. One can demonstrate this point by the following coordinates transformation,
    \begin{widetext}
    \be
   u=t\mp \int dr \left[\frac{kL(1-d  p) r^{2-p}}{2 (d +1) \left(d ^2 (2 d +1) p^3-p+2\right)} + C r^{\frac{2 p^2 d ^2}{p d -1}}\right]r^{\frac{pd(1+pd)}{1-pd}},
   \en
    \end{widetext}
   where we set $\cal G$=$C$. We note that if $\cal G$ is not a constant, we cannot obtain a metric like (\ref{metric1}) through the above transformation.
  Drawing an analogy with four-dimensional Vaidya metric \cite{vai}, we can regard the case with ``+" and $ \frac{ d{\cal G}}{du} >0$ in (\ref{dynametr}) as an absorbing star, and the case with ``$-$" and $ \frac{ d{\cal G}}{du} <0$ in (\ref{dynametr}) as a shining star in three-dimensional space-time, notwithstanding, the matter particles that they are absorbing or shining are not quanta of electromagnetic fields. We emphasize that three-dimensional Einstein gravity also permits non-trivial geometries in a sourced space-time, such as a charged BTZ \cite{cbtz}. So in principle, it also permits gravitational waves with sources \cite{QNM3}.

  The case $k=0$ is of special interests, which describes a circularly symmetric pure gravitational wave. In this case the situation with ``+" and $ \frac{ d{\cal G}}{du} >0$ still describes an absorbing star, while ``$-$" and $ \frac{ d{\cal G}}{du} <0$ for a shining star. However, there is nothing but gravity energy to absorbing or shining in this case: it is a solution for absorption or radiation of gravitational wave.   It realizes two impossibilities in the corresponding Einstein gravity: The first one is a circularly symmetric gravitational wave, and the second one is gravitational wave in three-dimensional vacuum. In this case the Ricci scalar $R$ vanishes.
  Gravity is the source of gravity, which is like the case of vacuum \sch~solution in 4-dimensions in some degree. The difference is that in the latter case Weyl tensor curves the space, while in the former case the Ricci tensor curves the space, for the Weyl tensor always vanishes in 3-dimensions.  The gravitational waves curve the space-time, which can be demonstrated by the contraction of Ricci tensor,
  \begin{widetext}
  \be
  R_{ab}R^{ab}=\pm\frac{(2 d-1) r^{-\frac{1+12 d^2}{d+2 d^2}} \left(\left(2 d-1-12 d^2+24 d^3\right) {\cal G}^2+4 d (1+2 d) r^{\frac{4 d}{1+2 d}}  \frac{ d{\cal G}}{du} \right)}{8 d^2 (1+2 d)^2}.
  \label{ricci}
  \en
    \end{widetext}
  One sees that a Birkhoff-like theorem  becomes invalid. The essential point is that a vanishing stress energy doe not imply a vanishing Ricci tensor in $f(R)$ gravity. From the above equation (\ref{ricci}), it is easy to see that the metric (\ref{dynametr}) comes back to a Minkowskian one when ${\cal G}(u)=0$, that is to say $\overline {A}=0$. So this dynamical space-time can be treated as a superposition of a gravitational wave $ \overline {A}du^2$ to a Minkowskian background ${\overline{B}(r)}dudr+r^2d\theta^2$ in the double null coordinates. Since $\cal G$ is an arbitrary function of $u$, this metric can describe any circularly symmetric gravitational wave in three-dimensional space-time. It can be tiny, as a fluctuation on the Minkowskian background. And it can also be large, as a nonperturbative gravitational wave. In any case it is an exact solution. So, in this work, we find a kind of proper ``object" to quantize. After quantization, they correspond to the on-shell gravitons. This may offer a useful clue to the gravity quantization in three-dimensional space-time.

   \section{Conclusion}
   In three-dimensional pure Einstein gravity, the local geometry is always trivial, therefore, gravitational wave can not propagate in such a space-time. This yields some confusions  of the corresponding quantum theory. It seems that the microscopic states in a BTZ black hole do not exist, since there is no wave in such a configuration. This also yields some confusions in the three-dimensional black hole thermodynamics. The quantum theory itself promise us some clues about this problem. A field, even in vacuum state, still shows some physical effects of itself, for example the Casimir effect. Classical fields with a zero stress energy has no effects on the gravitational field equation. However, when quantum effects are considered, they may yield higher derivative terms, like $f(R)$ theory, in the gravitational field equation\cite{bir}. Once the higher derivative terms are introduced, they will make significant and profound effects in three-dimensional space-time.

   First, as we have seen in this paper, non-trivial geometry and gravitational wave emerges naturally in higher derivative theories, which is radically different from those in the standard  Einstein gravity. This maybe helpful to solve the problem of the entropy of three-dimensional black hole. A theory with tiny non-zero $d$ may have enough microstates to yield the black hole entropy. It is not surprised that the quantum backreaction (resulting a higher derivative gravity) to illuminate this problem, since fundamentally the microscopic theory of black hole entropy is quantum theory. Second, to include the higher derivative terms is helpful to control the divergence in graviton scattering process, remarkably increasing the renormalizability of the theory. Thus it is worthy of studying the detailed properties of three-dimensional $f(R)$ gravity.  We find a circularly symmetric black hole solution with non-trivial geometries in $f(R)$ gravity. The spherically symmetric metrics can be derived directly from thermodynamic considerations \cite{zhang}, so we expect the black hole solution obtained here  may also have deep relations with thermodynamics and can be obtained in a similar manner. Besides, a dynamical solution is also obtained. When $k=\pm 1$ the solution describes an absorbing or shining star, depending on the sign of the stress energy. When $k=0$, it describes a dynamical space-time without source, which becomes a pure gravitational wave space-time. It is a circular wave, where the Birkhoff-like theorem is invalid. This solution can be treated as a wave propagating on a Minkowskian background.  Because of the arbitrary function ${\cal G}$ in this solution, in principle it can describe any wave with circularly symmetry. Therefore, the solution is a concrete example of the gravitational waves in three-dimensional space-time. After quantization, it may shed light on the microscopic origin of three-dimensional black hole, and further more, on the quantum theory of three-dimensional gravity.

 {\bf Acknowledgments.}
  H. Z. thanks S. Carlip for helpful discussions. This work is supported in part by the Program for Professor of Special Appointment (Eastern Scholar) at Shanghai Institutions of Higher Learning, National Education Foundation of China under grant No. 200931271104, and National Natural Science Foundation of China under Grant No. 11075106 and 11275128, and Shanghai Commission of Science and technology under Grant No.~12ZR1421700.


\begin{thebibliography}{99}

 \bibitem{carlip}
 S. Carlip,
Class.  Quant.  Grav. {\bf 22},  R85 (2005) [gr-qc/0503022].

 \bibitem{deser}
 S. Deser, R. Jackiw, and G. 't Hooft,
 Annals Phys. {\bf 152},
  220 (1984).


 \bibitem{btz}
 M.\ Ba\~nados, C.\ Teitelboim, and J.\ Zanelli,
 Phys.\ Rev.\ Lett.\ {\bf 69},  1849(1992).

\bibitem{Hawking}
S.~ Hawking, Nature {\bf 248},  30 (1974).

\bibitem{zhang}
H. Zhang, S. A. Hayward, X.-H. Zhai, X.-Z. Li, Phys.\ Rev.\ {\bf D89}, 064052 (2014) [arXiv:1304.3647 [physics.gen-ph]];
H. Zhang, Y. Hu and X. Z. Li, Phys. Rev. D 90, 024062 (2014) [arXiv:1406.0577 [gr-qc]];
 H.~Zhang and X.~Z.~Li,
  Phys.\ Lett.\ B {\bf 737} (2014) 395
  [arXiv:1406.1553 [gr-qc]].

   \bibitem{witten}
    E.~Witten,
  arXiv:0706.3359 [hep-th].

  \bibitem{deser2}
  S.~Deser, R.~Jackiw and S.~Templeton,
  Phys.\ Rev.\ Lett.\  {\bf 48}, 975 (1982).

  \bibitem{bht}
  E.~A.~Bergshoeff, O.~Hohm and P.~K.~Townsend,
  Phys.\ Rev.\ Lett.\  {\bf 102}, 201301 (2009)
  [arXiv:0901.1766 [hep-th]].

  \bibitem{bt2}
  E.~Bergshoeff, O.~Hohm, W.~Merbis, A.~J.~Routh and P.~K.~Townsend,
  arXiv:1404.2867 [hep-th]; E.~A.~Bergshoeff, O.~Hohm and P.~K.~Townsend,
  Phys.\ Rev.\ D {\bf 79}, 124042 (2009)
  [arXiv:0905.1259 [hep-th]]; Y.~Liu and Y.~-W.~Sun,
  Phys.\ Rev.\ D {\bf 79}, 126001 (2009)
  [arXiv:0904.0403 [hep-th]].

  \bibitem{ott}
  J.~Oliva, D.~Tempo and R.~Troncoso,
  JHEP {\bf 0907}, 011 (2009)
  [arXiv:0905.1545 [hep-th]].



  \bibitem{Blago}
  M.~Blagojevic and B.~Cvetkovic,
  Phys.\ Rev.\ D {\bf 85}, 104003 (2012)
  [arXiv:1201.4277 [gr-qc]].


  \bibitem{stelle}
 K. S. Stelle, Phys. Rev. {\bf D16}, 953 (1977).
   \bibitem{dewitt}
  R. Utiyama and B. De Witt, J. Math. Phys. \textbf{3}, 608 (1962); S. Deser, in {\it Gauge~Theories~and~Modern~Field~Theory},
  edited by R. Arnowit and P. Nath, MIT Press (1975).

 \bibitem{soti}
  T.~P.~Sotiriou and V.~Faraoni,
  Rev.\ Mod.\ Phys.\  {\bf 82}, 451 (2010)
  [arXiv:0805.1726 [gr-qc]].
\bibitem{vai}
  P. C. Vaidya, Proc. Indian Acad. Sci. \textbf{A33}, 264 (1951);
  R. W. Lindquist, R. A. Schwartz, and C. W. Misner, Phys. Rev. \textbf{137}, B1364 (1965).

 \bibitem{cbtz}
    M.\ Ba\~nados, M.\ Henneaux, C.\ Teitelboim, and J.\ Zanelli,
 Phys.\ Rev.\ \textbf{D48}, 1506 (1993)[gr-qc/9302012].

 \bibitem{QNM3}
  V.~Cardoso and J.~P.~S.~Lemos,
  Phys.\ Rev.\ D {\bf 63}, 124015 (2001)
  [gr-qc/0101052].


  \bibitem{bir}
 N. Birrell and P. Davies, {\it Quantum~Fields ~in~ Curved~ Space}, Cambrige University Press (1984).


\end{thebibliography}
\end{document}